\documentclass{pasj00}
\begin{document}
\SetRunningHead{J.\ Fukue}
{Relativistic Variable Eddington Factor}
\Received{yyyy/mm/dd}
\Accepted{yyyy/mm/dd}

\title{Relativistic Variable Eddington Factor}

\author{Jun \textsc{Fukue}} 
\affil{Astronomical Institute, Osaka Kyoiku University, 
Asahigaoka, Kashiwara, Osaka 582-8582}
\email{fukue@cc.osaka-kyoiku.ac.jp}



\maketitle


\begin{abstract}
We analytically derive a relativistic variable Eddington factor
in the relativistic radiative flow, 
and found that the Eddington factor depends on
the {\it velocity gradient} as well as the flow velocity.
When the gaseous flow is accelerated 
and there is a velocity gradient,
there also exists a density gradient.
As a result,
an unobstructed viewing range by a comoving observer,
where the optical depth measured from the comoving observer is unity,
is not a sphere, but becomes an oval shape elongated
in the direction of the flow;
we call it a {\it one-tau photo-oval}.
For the comoving observer,
an inner wall of the photo-oval
generally emits at a non-uniform intensity,
and has a relative velocity.
Thus, 
the comoving radiation fields observed by the comoving observer
becomes {\it anisotropic},
and the Eddington factor must deviate from
the value for the isotropic radiation fields.
In the case of a plane-parallel vertical flow,
we examine the photo-oval and obtain the Eddington factor.
In the sufficiently optically thick linear regime,
the Eddington factor is analytically expressed as
$f (\tau, \beta, \frac{d\beta}{d\tau}) 
= \frac{1}{3} ( 1 + \frac{16}{15} \frac{d\beta}{d\tau})$,
where $\tau$ is the optical depth and
$\beta$ ($=v/c$) is the flow speed normalized by the speed of light.
We also examine the linear and semi-linear regimes,
and found that the Eddington factor generally depends
both on the velocity and its gradient.
\end{abstract}

\section{Eddington Approximation}

In the moment formalism of radiation transfer in a static atmosphere
(Chandrasekhar 1960; Mihalas 1970; Rybicki, Lightman 1979;
Mihalas, Mihalas 1984, Shu 1991; Peraiah 2002; Castor 2004),
in order to close moment equations truncated at the finite order,
we usually adopt the {\it Eddington approximation}
in the inertial frame (laboratory frame) as a closure relation,
\begin{equation}
   P^{ij} = \frac{\delta^{ij}}{3} E,
\label{PE}
\end{equation}
where
$E$ and $P^{ij}$ are the radiation energy density and
the radiation stress in the inertial frame, respectively.

Similary,
in the moment formalim of radiation transfer in a relativistic flow
(Thorne 1981; Thorne et al. 1981; Flammang 1982, 1984;
Nobili et al. 1991, 1993; Park 2001, 2006; Takahashi 2007),
we often assume the Eddington approximation 
in the comoving frame (fluid frame),
\begin{equation}
   P_0^{ij} = \frac{\delta^{ij}}{3} E_0,
\label{P0E0}
\end{equation}
where
$E_0$ and $P_0^{ij}$ are the radiation energy density and
the radiation stress in the comoving frame, respectively.
Then, this closure relation in the comoving frame
is transformed to the relation in the inertial frame
(Hsieh, Spiegel 1976; Fukue et al. 1985;
Kato et al. 1998, 2008).

\begin{figure}
  \begin{center}
  \FigureFile(80mm,80mm){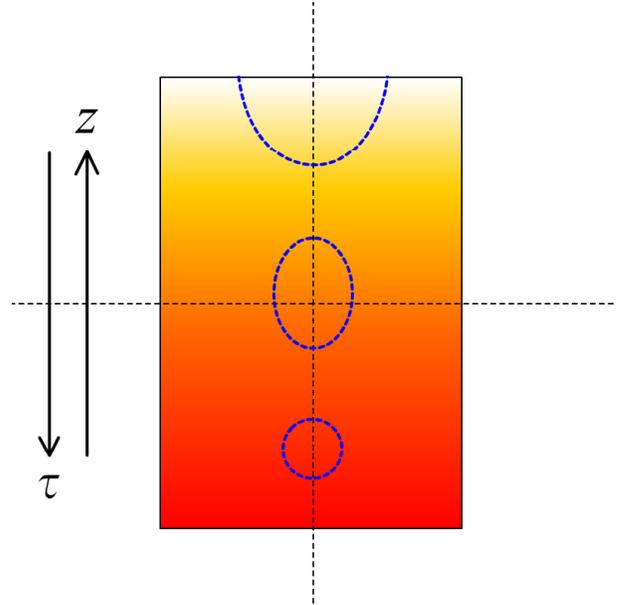}
  \end{center}
\caption{
Schematic picture of a relativistic radiative flow
in the vertical direction.
The flow is accelerated in the vertical ($z$) direction,
and has a velocity gradient.
The dashed curves are
one-tau photo-ovals observed by a comoving observer.
}
\end{figure}

As is well known, the Eddington approximation is valid
when the radiation fields are almost {\it isotropic}.
In the relativistically accelerating flow, however,
even for a comoving observer,
the comoving radiation fields become {\it anisotropic}.

Let us suppose a relativistic radiative flow,
which is accelerated in the vertical ($z$) direction,
and a comoving observer,
who moves upward with the flow,
at an optical depth $\tau$ (figure 1).

At the sufficiently deep inside the flow
(a dashed circle in figure 1),
where the optical depth is very large,
the mean free path of photons is very short.
In such a case,
within the range of the mean free path,
for the comoving observer
the flow is seen to be almost uniform;
the velocity gradient and resultant density gradient
can be negligible.
Hence,
the mean free path is the same in all directions,
or the shape of the one-tau range, to which the optical depth 
measured from the comoving observer is unity, is almost sphere;
we call it a {\it one-tau photo-sphere}.
As a result,
the comoving radiation fields is nearly isotropic,
and a usual Eddington approximation in the comoving frame is valid.

In the region
where the velocity is large and the density is low,
or
where the velocity gradient is large
in spite of the high density
(a dashed oval in figure 1),
within the range of the mean free path,
the velocity gradient cannot be neglected,
and the density is no longer uniform
even in the comoving frame of the gas.
Hence,
the mean free path becomes longer in the downstream direction
than in the upstream and other directions, and
the shape of the one-tau range elongates in the downstream direction;
we call it a {\it one-tau photo-oval}.
As a result,
the comoving radiation fields becomes {\it anisotropic},
and we should modify a usual Eddington approximation;
the Eddington factor may depend on 
the optical depth $\tau$,
the flow velocity $\beta$, and the velocity gradient $d\beta/d\tau$
in the relativistic regime.

In addition,
when the optical depth is sufficiently small
and/or the velocity gradient is suffiently large
(a dashed hemi-circle in figure 1),
the mean free path of photons in the downstream direction
become less than unity.
Hence,
the shape of the one-tau range is open in the downstream direction;
we call it a {\it one-tau photo-vessel}.

In order to obtain an appropriate form
of the relativistic variable Eddington factor $f(\tau, \beta, d\beta/d\tau)$,
we thus carefully treat and examine
the radiation fields in the comoving frame.
In this paper, at the first step, 
we examine the one-tau photo-oval
and derive the relativistic variable Eddington factor
in the linear and subrelativistic regimes.

In the next section
we examine the shape of one-tau photo ovals
seen by the comoving observer
under the several approximations,
including linear and semi-linear regimes.
In section 3
we analytically and numerically calculate
the comoving radiation fields within the photo-oval,
and derive the relativistic variable Eddington factor
under the several approximations.
In section 4,
we briefly discuss the importance of the velocity gradient
on the relativistic variable Eddington factor.
The final section is devoted to concluding remarks.


\section{One-Tau Photo-Oval Walls}

In this section
we first consider the shape of the one-tau photo-oval,
to which the optical depth measured from the comoving observer is unity,
in the vertical one-dimensional radiative flow
under the linear approximation.
We shall also obtain the breakup condition,
where a photo-oval is open toward the downstream direction
to become a photo-vessel (figure 1).

\begin{figure}
  \begin{center}
  \FigureFile(80mm,80mm){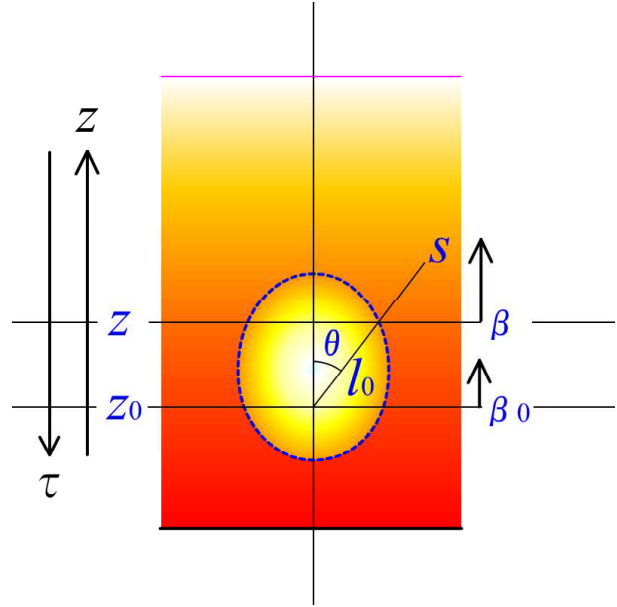}
  \end{center}
\caption{
One-tau photo-oval around a comoving observer
in the vertical ($z$) one-dimensional radiative flow.
The comoving observer is located at $z=z_0$,
where the flow speed is $\beta=\beta_0$.
In the $s$-direction,
where the angle measured from the downstream direction is $\theta$,
the mean free path is set to be $l_0$.
}
\end{figure}

The situation is schematically illustrated in figure 2.
In the vertical ($z$) flow,
we suppose a comoving observer located at $z=z_0$,
where the flow speed is $\beta=\beta_0$
($\beta$ is the speed normalized by the speed of light).
In the $s$-direction,
where the angle measured from the downstream direction is $\theta$,
the mean free path of photons is $l_0$.
The relation among these quantities is
\begin{equation}
   z - z_0 = s \cos\theta.
\label{zz0}
\end{equation}

The continuity equation for the present one-dimensional 
relativistic flow is
\begin{equation}
   \rho c\gamma\beta = J ~(={\rm const.}),
\label{rho}
\end{equation}
where $\rho$ is the proper gas density,
$\gamma$ ($=1/\sqrt{1-\beta^2}$) the Lorentz factor,
and $J$ the mass-flow rate per unit area.
In addition, the optical depth $\tau$ is defined by
\begin{equation}
   d\tau \equiv -\kappa \rho dz,
\label{tau}
\end{equation}
where $\kappa$ is the opacity,
which we assume constant in the present analysis.

In this paper
we use a linear approximation for the flow field;
that is to say, around the comoving observer
the flow speed is expanded as
\begin{equation}
   \beta = \beta_0 + \left.\frac{d\beta}{dz}\right|_0 ~(z-z_0),
\label{linear_beta}
\end{equation}
and we assume that the velocity gradient $d\beta/dz|_0$ is constant
in the one-tau photo oval range.

Under these situations and assumptions,
we examine three regimes below.

\subsection{Extremely Linear Regime}

In the linear regime,
the density is also expanded as
\begin{eqnarray}
   \rho &=& \rho_0 + \left.\frac{d\rho}{dz}\right|_0 ~(z-z_0)
\nonumber \\
        &=& \rho_0 + \left.\frac{d\rho}{d\beta}\frac{d\beta}{dz}\right|_0 ~(z-z_0)
\nonumber \\
        &=& \rho_0 - \left.\frac{\rho\gamma^2}{\beta}\frac{d\beta}{dz}\right|_0 ~(z-z_0),
\label{linear_rho}
\end{eqnarray}
where we use the continiuty equation (\ref{rho}),
and we also assume that the density gradient $d\rho/dz|_0$ is constant.

Then the optical depth $\tau_s$ along the $s$-direction
is easily calculated as
\begin{eqnarray}
   \tau_s &=& \int_0^{l_0} \kappa\rho ds
\nonumber \\
          &=& \int_0^{l_0} \kappa \left[ \rho_0 + \left.\frac{d\rho}{d\beta}\frac{d\beta}{dz}\right|_0 ~(z-z_0) \right] ds
\nonumber \\
          &=& \int_0^{l_0} \kappa \left[ \rho_0 + \left.\frac{d\rho}{d\beta}\frac{d\beta}{dz}\right|_0 ~s \cos\theta \right] ds
\nonumber \\
          &=& \kappa\rho_0 l_0 + \kappa \left.\frac{d\rho}{d\beta}\frac{d\beta}{dz}\right|_0 \frac{\cos\theta}{2} l_0^2.
\label{linear_tau_s}
\end{eqnarray}
Hence, the length $l_0$ of the one-tau range ($\tau_s=1$) is determined by
the quadratic equation,
\begin{equation}
          \kappa \left.\frac{d\rho}{d\beta}\frac{d\beta}{dz}\right|_0 \frac{\cos\theta}{2} l_0^2 + \kappa\rho_0 l_0 -1 = 0.
\label{linear_l0}
\end{equation}

As is easily seen from equation (\ref{linear_l0}),
if there is no velocity gradient,
$\kappa\rho_0 l_0=1$, and
the shape of the one-tau range is a sphere
(one-tau photo-sphere).

If the deviation from the sphere is sufficiently small
($\kappa\rho_0 l_0 \sim 1$),
the shape is approximately calculated as a correction factor,
\begin{eqnarray}
   \kappa\rho_0 l_0 &=& 1 - \kappa \left.\frac{d\rho}{d\beta}\frac{d\beta}{dz}\right|_0 \frac{\cos\theta}{2} l_0^2
\nonumber \\
                    &\sim& 1 - \kappa \left.\frac{d\rho}{d\beta}\frac{d\beta}{dz}\right|_0 \frac{\cos\theta}{2} \frac{1}{\kappa^2\rho_0^2}
\nonumber \\
                    &=& 1 - \frac{1}{\kappa\rho_0^2} \left.\frac{d\rho}{d\beta}\frac{d\beta}{dz}\right|_0 \frac{\cos\theta}{2} 
\nonumber \\
                    &=& 1 + \frac{1}{\rho_0} \left.\frac{d\rho}{d\beta}\frac{d\beta}{d\tau}\right|_0 \frac{\cos\theta}{2},
\nonumber \\
                    &=& 1 - \left.\frac{\gamma^2}{\beta}\frac{d\beta}{d\tau}\right|_0 \frac{\cos\theta}{2}.
\label{linear_l01}
\end{eqnarray}
Thus, a small correction is proportional to $\cos\theta$,
and one-tau oval is elongated toward the downstream direction.

\subsection{Linear Regime}

In general the quadratic equation (\ref{linear_l0}) can be easily solved.
The quadratic equation (\ref{linear_l0}) is rearranged as
\begin{equation}
      a \frac{\cos\theta}{2} (\kappa\rho_0 l_0)^2 + \kappa\rho_0 l_0 -1 = 0,
\label{linear_l02}
\end{equation}
and the required solution is
\begin{equation}
     \kappa \rho_0 l_0 = \frac{1-\sqrt{1-2a\cos\theta}}{a\cos\theta},
\label{linear_kapparhol}
\end{equation}
where
\begin{eqnarray}
    a &\equiv& - \frac{1}{\kappa\rho_0^2} \left.\frac{d\rho}{d\beta}\frac{d\beta}{dz}\right|_0
\nonumber \\
      &=& + \frac{1}{\rho_0} \left.\frac{d\rho}{d\beta}\frac{d\beta}{d\tau}\right|_0
\nonumber \\
      &=& - \left.\frac{\gamma^2}{\beta}\frac{d\beta}{d\tau}\right|_0.
\label{linear_a}
\end{eqnarray}

\begin{figure}
  \begin{center}
  \FigureFile(80mm,80mm){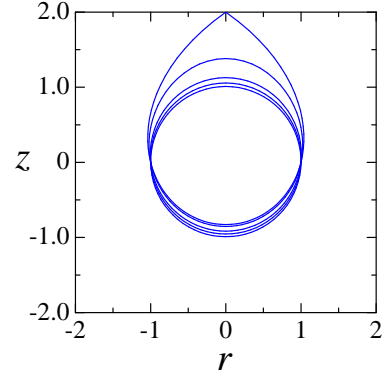}
  \end{center}
\caption{
Shapes of one-tau photo-ovals in the linear regime
for several values of parameter $a$.
The values of parameter $a$ are 0.02, 0.1, 0.2, 0.4, and 0.5
 from spherical to elongated.
}
\end{figure}

In figure 3
the solutions (\ref{linear_kapparhol}) 
of the quadratic equation (\ref{linear_l02})
are shown for several values of parameter $a$.
The values of parameter $a$ are 0.02, 0.1, 0.2, 0.4, and 0.5
 from spherical to elongated.
For $a>0.5$,
in the downstream direction
the solution disappears and the photo-oval
becomes open to be a photo-vessel.

 From the solution (\ref{linear_kapparhol}),
the breakup condition, where the photo-oval is open
in the downstream direction, is given by
$1-2a\cos\theta<0$, or
\begin{equation}
      - \left.\frac{d\beta}{d\tau}\right|_0 \cos\theta
      > \left.\frac{\beta}{2\gamma^2}\right|_0.
\label{linear_breakup}
\end{equation}
This breakup condition is shown in figure 4.
Below the curve photo-ovals are closed,
whereas above the curve they are open
in the downstream direction
and the present approximation is violated.

\begin{figure}
  \begin{center}
  \FigureFile(80mm,80mm){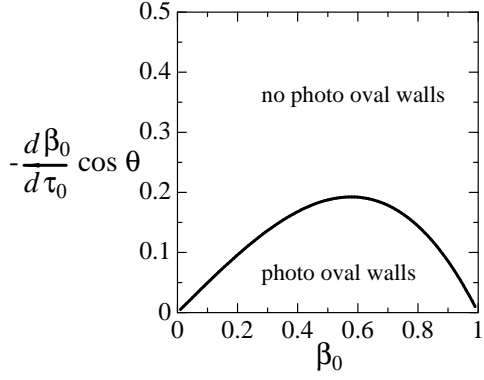}
  \end{center}
\caption{
Breakup condition for photo-ovals in the linear regime.
Below the curve photo-ovals are closed,
whereas above the curve they are open
in the downstream direction.
}
\end{figure}

Using the continuity equation (\ref{rho}) and the optical depth (\ref{tau}),
the breakup condition (\ref{linear_breakup}) is also written as
\begin{equation}
      \left.\frac{d\beta}{dz}\right|_0 \cos\theta
      > \frac{\kappa J}{2c} \left.\frac{1}{\gamma^3}\right|_0.
\label{linear_breakup2}
\end{equation}
Hence,
the condition is somewhat different appearance,
whether we consider the velocity gradient
in the linear length $z$ or the optical depth $\tau$.

In any case,
at this stage we demonstrate that
both the shape of the one-tau photo oval and its breakup condition
depend on the flow {\it velocity} and the {\it velocity gradient}.

\subsection{Semi-Linear Regime}

When the flow speed is high and the density decreases,
the mean free path of photons becomes large,
and the one-tau range would expand.
Then,
even if the flow field around the comoving observer
is assumed to be linear in a manner of equation (\ref{linear_beta}),
there is no guarantee for the linearity of the density distribution,
and we should determine the density 
by continuity equation (\ref{rho}).
We here examine such a semi-linear regime.

Using continuity equation (\ref{rho}),
the optical depth $\tau_s$ along the $s$-direction
is expressed as
\begin{eqnarray}
   \tau_s &=& \int_0^{l_0} \kappa\rho ds
\nonumber \\
    &=& \frac{\kappa J}{c} \int_0^{l_0} \frac{\sqrt{1-\beta^2}}{\beta} ds.
\label{semi_tau_s}
\end{eqnarray}
 From equations (\ref{zz0}) and (\ref{linear_beta}), we have
\begin{eqnarray}
   \beta &=& \beta_0 + \left.\frac{d\beta}{dz}\right|_0 ~s \cos\theta,
\\
   d\beta &=& \left.\frac{d\beta}{dz}\right|_0 ~\cos\theta ~ds,
\label{semi_beta}
\end{eqnarray}
and the integral is transformed as
\begin{equation}
   \tau_s = \frac{\kappa J}{c} \frac{1}{\left.\frac{\displaystyle d\beta}{\displaystyle dz}\right|_0 \cos\theta}
              \int_{\beta_0}^{\beta} \frac{\sqrt{1-\beta^2}}{\beta} d\beta,
\label{semi_tau_s2}
\end{equation}
where we use the assumption that the velocity gradient is constant.
This equation (\ref{semi_tau_s2}) is analytically integrated to give
\begin{eqnarray}
   \tau_s &=& \frac{\kappa J}{c} \frac{1}{\left.\frac{\displaystyle d\beta}{\displaystyle dz}\right|_0 \cos\theta}
              \left[
                 \sqrt{1-\beta^2} - \sqrt{1-\beta_0^2}
              \right.
\nonumber \\
          &&  \left.
~~~~~~~~~~~~~~~~~~~~
                 + \log \frac{\beta}{\beta_0}
                        \frac{1+\sqrt{1-\beta_0^2}}{1+\sqrt{1-\beta^2}}
              \right]
\nonumber \\
          &=& - \frac{1}{\left.\frac{\displaystyle d\beta}{\displaystyle d\tau}\right|_0 \cos\theta}
                \gamma_0 \beta_0
              \left[
                 \sqrt{1-\beta^2} - \sqrt{1-\beta_0^2}
              \right.
\nonumber \\
          &&  \left.
~~~~~~~~~~~~~~~~~~~~
                 + \log \frac{\beta}{\beta_0}
                        \frac{1+\sqrt{1-\beta_0^2}}{1+\sqrt{1-\beta^2}}
              \right].
\label{semi_tau_s3}
\end{eqnarray}

\begin{figure}
  \begin{center}
  \FigureFile(80mm,80mm){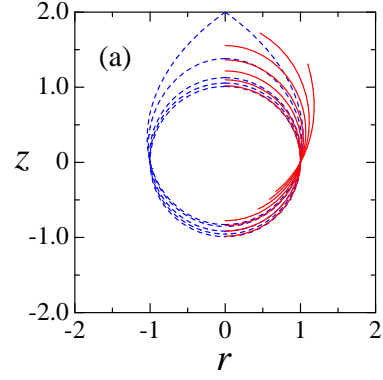}
  \FigureFile(80mm,80mm){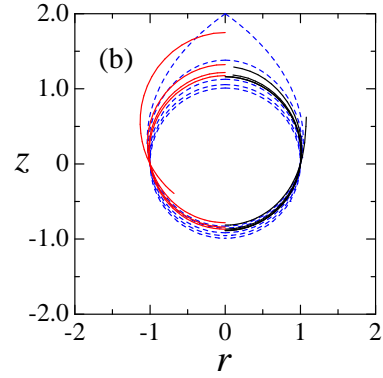}
  \end{center}
\caption{
Shapes of one-tau photo-ovals in the semi-linear regime
for several values of $\beta_0$ and $-d\beta/d\tau|_0$.
The values of parameters are
(a) $\beta_0=0.3$ and $-d\beta/d\tau|_0=$
0.01, 0.05, 0.10, 0.15, 0.20, 0.25, and 0.30
 from spherical to elongated, and
(b) $-d\beta/d\tau|_0 =0.1$ and
$\beta_0=$ 0.1, 0.2, 0.3, and 0.4
 from elongated to spherical in the leftside,
while 0.5, 0.6, 0.7, and 0.9
 from spherical to elongated in the rightside.
The dashed curves are photo-ovals in the linear regime
shown in figure 3.
}
\end{figure}

Thus, the length $l_0$ of the one-tau range ($\tau_s=1$) 
is finally determined,
with the help of equation (\ref{semi_beta})
and the definition of the optical depth (\ref{tau}),
by the following equations:
\begin{eqnarray}
   0 &=& - {\left.\frac{d\beta}{d\tau}\right|_0 \cos\theta}
              -  \frac{\beta_0}{\sqrt{1-\beta_0^2}}
              \left[
                 \sqrt{1-\beta^2} - \sqrt{1-\beta_0^2}
              \right.
\nonumber \\
          &&  \left.
~~~~~~~~~~~~~~~~~~~~
                 + \log \frac{\beta}{\beta_0}
                        \frac{1+\sqrt{1-\beta_0^2}}{1+\sqrt{1-\beta^2}}
              \right],
\label{semi_l0}
\\
    \beta &=& \beta_0 - {\left.\frac{d\beta}{d\tau}\right|_0 \cos\theta}
                  ~\kappa\rho_0 l_0.
\label{semi_beta0}
\end{eqnarray}

In figure 5
the numerical solutions of equations (\ref{semi_l0}) and (\ref{semi_beta0})
are shown for several values of $\beta_0$ and $-d\beta/d\tau|_0$;
i.e.,
the one-tau length $\kappa\rho_0 l_0$
is plotted as a function of $\theta$.
In figure 5a
the values of parameters are
$\beta_0=0.3$ and $-d\beta/d\tau|_0=$
0.01, 0.05, 0.10, 0.15, 0.20, 0.25, and 0.30
 from spherical to elongated.
When the velocity gradient is small,
a photo-oval is spherical.
As the velocity gradient increases,
it elongates in the downstream direction,
and breaks up in the downstream direction
at sufficiently large gradients.
In addition,
the shape of photo-ovals are somewhat fat,
compared with those in the linear regime
shown by dashed curves.

In figure 5b, on the other hand,
the values of parameters are
$-d\beta/d\tau|_0 =0.1$ and
$\beta_0=$ 0.1, 0.2, 0.3, and 0.4
 from elongated to spherical in the leftside,
while 0.5, 0.6, 0.7, and 0.9
 from spherical to elongated in the rightside.
As is seen in figure 5b,
a photo-oval elongates in the case of smaller $\beta_0$.
This is because, for the fixed $-d\beta/d\tau|_0$,
the velocity difference from the comoving observer
becomes relatively large,
if the velocity is small.

We can further obtain the breakup condition,
where the photo-oval is open in the downstream direction.
Under the present approximation,
equation (\ref{semi_l0}) has no solution,
when $\beta=1$.
Or,
the breakup condition is
\begin{equation}
   - {\left.\frac{d\beta}{d\tau}\right|_0 \cos\theta} >
              \beta_0
              \left[
                  \frac{1}{\sqrt{1-\beta_0^2}}
                  \log \frac{1+\sqrt{1-\beta_0^2}}{\beta_0} -1
              \right].
\label{semi_break}
\end{equation}
This breakup condition is shown in figure 6.
Below the solid curve photo-ovals are closed,
whereas above the curve they are open
in the downstream direction.

\begin{figure}
  \begin{center}
  \FigureFile(80mm,80mm){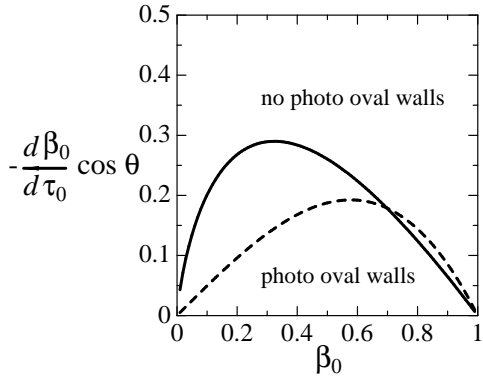}
  \end{center}
\caption{
Breakup condition for photo-ovals in the semi-linear regime.
Below the solid curve photo-ovals are closed,
whereas above the curve they are open
in the downstream direction.
The dashed curve is the breakup condition for the linear regime
shown in figure 4.
}
\end{figure}


\section{Comoving Radiation Fields}

We now examine the radiation fields in the comoving frame;
the radiation emitted from the one-tau photo oval walls
forms the comoving radiation fields
at the position of the comoving observer (figure 7).

\begin{figure}
  \begin{center}
  \FigureFile(80mm,80mm){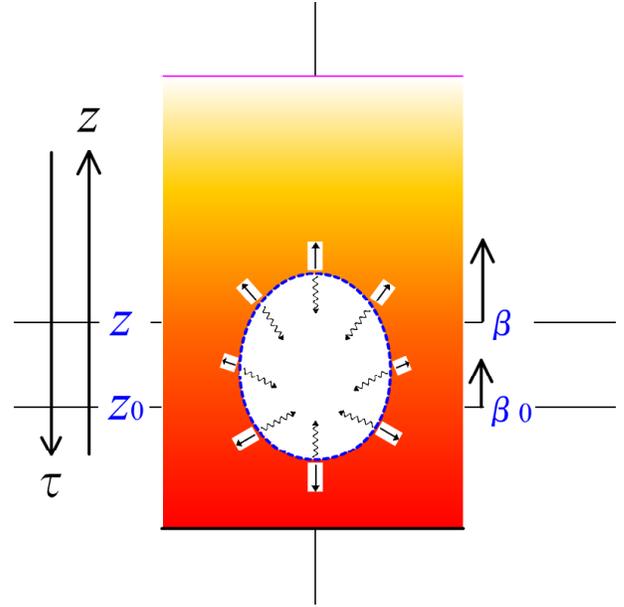}
  \end{center}
\caption{
Radiation fields in the one-tau photo-oval around a comoving observer
in the vertical one-dimensional radiative flow.
The radiation fields become anisotropic,
since in general the emitted intensity is not uniform and
it is redshifted due to the velocity difference.
}
\end{figure}

In the static and optically thick atmosphere,
the radiation fields are isotropic and uniform.
In the present moving atmosphere, on the other hand,
there are two factors, non-uniform intensity and redshift,
which make radiation fields anisotropic.

First, the radiative intensity $I_{\rm co}$ in the comoving frame
emitted from the one-tau photo-oval walls is not generally constant,
but is a function of $z$ (or $\tau$) and $\mu$ ($=\cos\theta$),
due to the gradient of physical quantities.
If the optical depth is sufficiently large,
the $\mu$-dependence could be safely ignored,
and the emitted intensity would be assumed to be isotropic.
In general, however,
the radiative intensity increases with the optical depth,
and therefore,
the intensity from the upstream direction is slightly
larger than that from the downstream direction
in the present one-dimensional vertical flow.
Hence, the effect of non-uniformity of intensity
generally acts as a force to accelerate a comoving observer.

In this paper
we use a linear approximation;
that is, around the comoving observer
both the flow speed and intensity in the comoving frame are expanded as
\begin{eqnarray}
   \beta &=& \beta_0 + \left.\frac{d\beta}{dz}\right|_0 ~(z-z_0)
          =  \beta_0 - {\left.\frac{d\beta}{d\tau}\right|_0 \cos\theta}
                  ~\kappa\rho_0 l_0,
\label{RF_beta}
\\
   I     &=& I_0 + \left.\frac{dI}{dz}\right|_0 ~(z-z_0)
          =  I_0 - {\left.\frac{dI}{d\tau}\right|_0 \cos\theta}
                  ~\kappa\rho_0 l_0,
\label{RF_beta}
\end{eqnarray}
where we use equations (\ref{zz0}) and (\ref{tau}),
and we assume that the intensity gradient $dI/dz|_0$ is constant.

Second,
the intensity observed by the comoving observer
is redshifted (Doppler shifted) due to velocity difference between
the comoving observer and the one-tau photo-oval walls.
In the accelerating flow,
where the flow speed increases toward the downstream direction,
the relative velocity is generally positive (figure 7)
except for some special direction of $\theta=\pi/2$.
Hence, the Doppler shift of intensity also causes anisotropy
of the radiation fields at the comoving observer.

The relative speed $\Delta \beta$ between the comoving observer
and the one-tau photo-oval walls is given
by the relativistic summation law, but
in the linear regime we approximate the usual form as
\begin{equation}
   \Delta \beta = \frac{\beta - \beta_0}{1-\beta\beta_0}
                \sim \beta - \beta_0
          =  - {\left.\frac{d\beta}{d\tau}\right|_0 \cos\theta}
                  ~\kappa\rho_0 l_0.
\end{equation}
Using this relative speed,
the redshift $z$ is expressed as
\begin{equation}
   1+z = \frac{1+\Delta \beta \cos\theta}{\sqrt{1-(\Delta\beta)^2}}.
\end{equation}
Thus, the observed intensity $I_{\rm co}$ of the comoving observer
becomes
\begin{eqnarray}
   I_{\rm co} &=& \frac{I}{(1+z)^4}
\nonumber \\
              &=& \left( I_0 - {\left.\frac{dI}{d\tau}\right|_0 \cos\theta}
                  ~\kappa\rho_0 l_0 \right)
\nonumber \\
          && \times \frac{ \left[ 1 - \left( {\left.\frac{\displaystyle d\beta}{\displaystyle d\tau}\right|_0 \cos\theta}~\kappa\rho_0 l_0 \right)^2 \right]^2 }
                     { \left( 1 - {\left.\frac{\displaystyle d\beta}{\displaystyle d\tau}\right|_0 \cos^2 \theta}~\kappa\rho_0 l_0 \right)^4 }.
\label{RF_Ico}
\end{eqnarray}

\subsection{Extremely Linear Regime}

Again, we first consider the sufficiently linear regime,
where the velocity gradient is sufficiently small.
In this case, the observed intensity (\ref{RF_Ico})
is linearly expanded as
\begin{eqnarray}
   I_{\rm co} &\sim& 
    \left( I_0 - {\left.\frac{dI}{d\tau}\right|_0 \cos\theta}~\kappa\rho_0 l_0 \right)
 \left( 1 + 4 {\left.\frac{d\beta}{d\tau}\right|_0 \cos^2 \theta}~\kappa\rho_0 l_0 \right)
\nonumber \\
    & \sim & 
   I_0 \left( 1 - \frac{1}{I_0}{\left.\frac{dI}{d\tau}\right|_0 \cos\theta} + 4 {\left.\frac{d\beta}{d\tau}\right|_0 \cos^2 \theta} \right),
\label{RF_Ico_linear}
\end{eqnarray}
where we use equation (\ref{linear_l01}) for $\kappa\rho_0 l_0$.

In this extremely linear regime,
the radiation energy density $E_{\rm co}$,
the radiative flux $F_{\rm co}$, and
the radiation pressure $P_{\rm co}$
measured by the comoving observer is analytically calculated as
\begin{eqnarray}
   cE_{\rm co} &\equiv& \int I_{\rm co} d\Omega_{\rm co}
\nonumber \\
           &\sim& 2\pi I_0 \int_0^\pi
\left( 1 - \frac{1}{I_0}{\left.\frac{dI}{d\tau}\right|_0 \cos\theta} + 4 {\left.\frac{d\beta}{d\tau}\right|_0 \cos^2 \theta} \right) \sin\theta d\theta
\nonumber \\
           &=& 4\pi I_0 \left( 1 + \left.\frac{4}{3}\frac{d\beta}{d\tau}\right|_0 \right),
\label{RF_Eco_linear}
\\
   F_{\rm co} &\equiv& \int I_{\rm co} \cos\theta d\Omega_{\rm co}
\nonumber \\
           &=& 4\pi I_0 \frac{1}{3} \left( -\left.\frac{dI}{Id\tau}\right|_0 \right),
\label{RF_Fco_linear}
\\
   cP_{\rm co} &\equiv& \int I_{\rm co} \cos^2\theta d\Omega_{\rm co}
\nonumber \\
           &=& 4\pi I_0 \frac{1}{3} \left( 1 + \left.\frac{12}{5}\frac{d\beta}{d\tau}\right|_0 \right).
\label{RF_Pco_linear}
\end{eqnarray}

As already stated,
if the emitted intensity from the one-tau photo-oval walls
is not uniform, and has a gradient toward the downstream direction,
as shown in equation (\ref{RF_Ico_linear}),
this intensity gradient acts as a force to push
the comoving observer toward the downstream direction,
as shown in equation (\ref{RF_Fco_linear}).
It should be noted that the radiative force
received by the comoving observer is
$-\kappa F_{\rm co}/c$.
However, this intensity gradient does not affect
the radiation energy density and the radiation pressure.

On the other hand, in this liear regime
the relative velocity and resultant redshift is always positive 
in any direction except for $\theta=\pi/2$,
and therefore, the observed intensity is slightly reduced
in all direction, as shown in equation (\ref{RF_Ico_linear}).
This velocity gradient affects
the radiation energy density and the radiation pressure,
as shown in equations (\ref{RF_Eco_linear}) and (\ref{RF_Pco_linear}),
but the radiative flux.
That is,
due to the velocity gradient,
both the radiation energy density and the radiation pressure
slightly decrease, compared with the non-relativistic case.

As a result, in this linear regime
the Eddinton factor in the comoving frame becomes
\begin{equation}
   f \equiv \frac{P_{\rm co}}{E_{\rm co}}
     \sim \frac{1}{3}
   \left( 1 + \left.\frac{16}{15}\frac{d\beta}{d\tau}\right|_0 \right).
\label{f_linear}
\end{equation}
Hence,
in the sufficiently linear regime,
where the velocity gradient is sufficiently small,
the relativistic variable Eddington factor depends only 
on the velocity gradient
to the flow optical depth.

It should be noted that equation (\ref{f_linear})
is also expressed as
\begin{equation}
   f \sim \frac{1}{3}
   \left( 1 - \left.\frac{16}{15}\frac{d\beta}{\kappa \rho dz}\right|_0 \right)
     \sim \frac{1}{3}
   \left( 1 - \left.\frac{16}{15}\frac{c\gamma\beta}{\kappa J}\frac{d\beta}{dz}\right|_0 \right).
\label{f_linear2}
\end{equation}
Hence, in this sense
the relativistic variable Eddington factor depend
both on the velocity and its gradient to the real length.

As is seen in equations (\ref{f_linear}) and (\ref{f_linear2}),
the Eddington factor slightly decreases
for the subrelativsitic flow with the velocity gradient.

\subsection{Linear Regime}

In the linear regime,
using the solutions (\ref{linear_kapparhol}),
we can numerically integrate the comoving intensity (\ref{RF_Ico_linear})
to obtain the comoving radiation fields.
Since the intensity gradient does not affect the Eddington factor
in the present linear treatment,
we shall drop the intensity gradient in what follows.
Hence, the parameters are
the velocity $\beta_0$ and its gradient $-d\beta/d\tau|_0$
of the comoving observer.
In addition, the integration in the polar direction
is restricted in the range of 
$\theta_1 = \cos^{-1}(1/2a) \leq \theta \leq \pi$,
although we do not consider a photo-vessel in this paper.

\begin{figure}
  \begin{center}
  \FigureFile(80mm,80mm){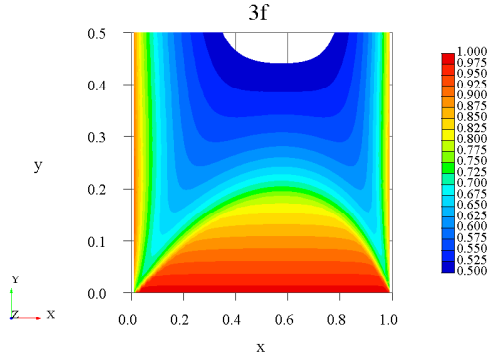}
  \end{center}
\caption{
Relativistic variable Eddington factor $f(\beta, -d\beta/d\tau)$
in the linear regime.
The value of $3f$ is plotted by a contour map
as a function of the velocity $\beta_0$ (abscissa)
and the velocity gradient $-d\beta/d\tau|_0$ (ordinate).
Reflecting the breakup condition in the linear regime,
the value of $3f$ quickly drops
at the outer edge of the one-tau photo-oval.
}
\end{figure}

The numerical result is shown in figure 8.
In figure 8, $3f$,
the three times of the Eddington factor $f(\beta, -d\beta/d\tau)$,
is plotted by a contour map
as functions of the velocity $\beta_0$ (abscissa)
and the velocity gradient $-d\beta/d\tau|_0$ (ordinate).
Reflecting the breakup condition in the linear regime
(see figure 4),
the value of $3f$ quickly drops
at the outer edge of the one-tau photo-oval.
Within the one-tau photo-oval region, 
the relativistic variable Eddington factor is well
fitted by equation (\ref{f_linear}),
except for the boundary region.

\subsection{Semi-Linear Regime}

Similary, in the semi-linear regime,
using equations (\ref{semi_l0}) and (\ref{semi_beta0}),
we can numerically integrate the comoving intensity (\ref{RF_Ico_linear})
to obtain the comoving radiation fields.
The parameters are
the velocity $\beta_0$ and its gradient $-d\beta/d\tau|_0$
of the comoving observer.
In addition, the integration in the polar direction
is restricted in the range of 
$\theta_1 = \leq \theta \leq \pi$,
where $\theta_1$ is given by equation (\ref{semi_break}).

\begin{figure}
  \begin{center}
  \FigureFile(80mm,80mm){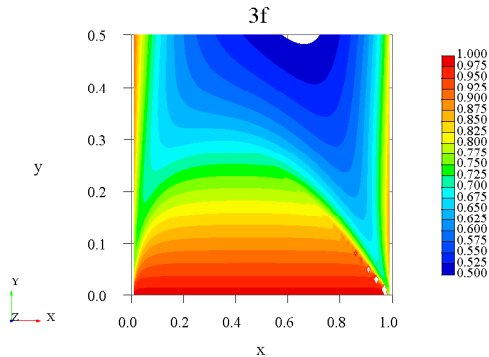}
  \end{center}
\caption{
Relativistic variable Eddington factor $f(\beta, -d\beta/d\tau)$
in the semi-linear regime.
The value of $3f$ is plotted by a contour map
as functions of the velocity $\beta_0$ (abscissa)
and the velocity gradient $-d\beta/d\tau|_0$ (ordinate).
Reflecting the breakup condition in the linear regime,
the value of $3f$ quickly drops
at the outer edge of the one-tau photo-oval.
}
\end{figure}

The numerical result is shown in figure 9.
In figure 9, $3f$,
the three times of the Eddington factor $f(\beta, -d\beta/d\tau)$,
is plotted by a contour map
as functions of the velocity $\beta_0$ (abscissa)
and the velocity gradient $-d\beta/d\tau|_0$ (ordinate).
Reflecting the breakup condition in the linear regime
(see figure 6),
the value of $3f$ quickly drops
at the outer edge of the one-tau photo-oval.
Within the one-tau photo-oval region, 
the relativistic variable Eddington factor is well
fitted by equation (\ref{f_linear}),
except for the boundary region.

\section{Discussion}

Moment equations for relativistic radiation transfer
have been derived in several literature
(Lindquist 1966; Anderson, Spiegel 1972; Hsieh, Spiegel 1976;
Thorne 1981; Thorne et al. 1981; Udey, Israel 1982; Schweizer 1982;
Flammang 1982, 1984; Nobili et al. 1991, 1993; Park 2001, 2006
for spherically symmetric problems;
Takahashi 2007 for the Kerr metric).
A complete set of moment equations for relativistic flow
is given by the projected, symmetric, trace-free formalism
by Thorne (1981).
Since a moment expansion gives an infinite set of equations,
one must truncate the expansion at the finite order
by adopting a suitable closure assumption,
in order to make the transfer problem tractable.
When one truncate at the second order, for example,
the Eddington approximation (\ref{P0E0}) has been usually adopted
as a closure relation.

Such radiation moment formalism is quite convenient,
and it is a powerful tool for tackling problems
of relativistic radiation hydrodynamics.
However, its validity is never known
unless fully angle-dependent radiation transfer equation
is solved.

Actually, the pathological behavior
in relativistic radiation moment equations
has been pointed out and examined
(Turolla, Nobili 1988; Nobili et al. 1991; Turolla et al. 1995;
Dullemond 1999).
Namely, the moment equations for radiation transfer
in relativistically moving media
can generally have singular (critical) points.
As a result, solutions behave pathologically
in regions of strong velocity gradients.
The appearance of singularities is explained
because we approximate the full transfer equations
with a finite number of moments (Dullemond 1999).
For example, in the one-dimensional relativistic radiation flow
using the Eddington approximation (\ref{P0E0}),
where the moment equations are truncated at the second order,
the singularity appears when the flow velocity becomes
$\beta=v/c=\pm 1/\sqrt{3}$.
Hence, under the traditional Eddington approximation,
we cannot obtain solutions
accelerated beyond $c/\sqrt{3}$, 
altough there exists a decelerating solution (Fukue 2005).

As already discussed in the introduction,
the invalidity of the Eddington approximation
in such a relativistic flow can be understood as follows.
In adopting the Eddington approximation (\ref{P0E0}), 
we assume that within the mean free path of photons
the radiation field is {\it isotropic} in the comoving frame.
However, in the relativistic regime,
where the velocity gradient becomes large
and there exist the Doppler and aberration effects of photons,
the isotropy of the radiation field may break down
even in the comoving frame.

Roughly speaking, the photon mean-free path $\ell$ in the comoving frame is 
$\ell \sim 1/(\kappa \rho)$,
where $\kappa$ is the opacity measured in the comoving frame
and $\rho$ is the proper density.
When there exists a velocity gradient, $dv/dz$,
the velocity increase at the distance of $\ell$
is estimated as
\begin{equation}
   \Delta v = \ell \frac{dv}{dz} = \frac{dv}{\kappa \rho dz}
   = -\frac{dv}{d\tau}.
\end{equation}
In order for the radiation fields to be isotropic in the comoving frame,
this velocity increase should be sufficiently smaller than
the speed of light.
Otherwise, the usual Eddington approximation in the comoving frame
would be violated.
In addition, photons suffer from
the Doppler and aberration effects,
which change the radiation fields in the comoving frame.
In such a case, we should modify the closure relation
in the case of subrelativistic to relativistic regimes,
as in the case of optically thick to thin regimes.


In order to improve the situation we are confronted with,
instead of the usual Eddington approximation,
we can adopt a {\it variable Eddington factor},
which depends on the flow velocity $\beta$ ($=v/c$)
and the velocity gradient $d\beta/d\tau$
 as well as the optical depth $\tau$
(Fukue 2006; Fukue, Akizuki 2006a, b; Akizuki, Fukue 2007;
Koizumi, Umemura 2007).
That is,
in one-dimensional flows
the variable Eddington factor $f(\tau, \beta, d\beta/d\tau)$ 
is generally defined as
\begin{equation}
   P_{\rm co} = f(\tau, \beta, d\beta/d\tau) E_{\rm co},
\label{PcoEco}
\end{equation}
where $E_{\rm co}$ and $P_{\rm co}$ are the radiation energy density
and the radiation stress tensor in the comoving frame, respectively.
The closure relation in the inertial frame
for one-dimensional flows then becomes
\begin{equation}
   cP(1-f\beta^2) = cE(f-\beta^2) + 2F\beta(1-f),
\label{PEF}
\end{equation}
where $E$, $F$, and $P$ are the radiation energy density,
the radiative flux, and the radiation pressure
in the inertial frame, respectively
(Kato et al. 1998, 2008).

The function $f(\tau, \beta, d\beta/d\tau)$ 
must reduce to 1/3 in the non-relativistic limit of $\beta=0$
or in the uniform flow of $d\beta/d\tau=0$, 
whereas it would approach unity
in the extremely relativistic limit of $\beta=1$.
Furthermore, in the sufficiently optically thick regime
this function approaches 1/3,
while in the optically thin limit it reduces to an appropriate form
determined by the geometry under the considerations.
In the plane-parallel case, for instance,
the variable Eddinton factor in the optically thin limit
is analytically derived as
$$
   f = \frac{1 - 3\beta + 3\beta^2}{3 - 3\beta + \beta^2},
$$
since $cE = 2F = 3cP$ in equation (\ref{PEF}) in the optically thin limit
(see also Kato et al. 2008; Koizumi, Umemura 2007).

In these recent studies
(Fukue 2006; Fukue, Akizuki 2006a, b; Akizuki, Fukue 2007;
Koizumi, Umemura 2007),
they considered mainly
the dependence of the variable Eddington factor
on the flow velocity.
Under the discussion above,
in this paper
we consider the dependence of the variabel Eddington factor
on the velocity gradient,
using the approach of the comoving observer's view point.

As a result, for example,
we derived the form of the relativistic variable Eddington factor
(\ref{f_linear}) and (\ref{f_linear2}),
where
the Eddington factor slightly decreases
for the subrelativsitic flow with the velocity gradient.
In the extremely linear regime,
the comoving radiation fields including
$P^{xx}$ and $P^{yy}$ perpendicular to the flow, and $P^{zz}$,
are calculated as
\begin{eqnarray}
   cE_{\rm co} &\sim& 4\pi I_0 \left( 1 + \left.\frac{4}{3}\frac{d\beta}{d\tau}\right|_0 \right),
\label{RF_Eco}
\\
   cP_{\rm co}^{xx} &\sim& 4\pi I_0 \frac{1}{3} \left( 1 + \left.\frac{4}{5}\frac{d\beta}{d\tau}\right|_0 \right),
\label{RF_Pcoxx}
\\
   cP_{\rm co}^{yy} &\sim& 4\pi I_0 \frac{1}{3} \left( 1 + \left.\frac{4}{5}\frac{d\beta}{d\tau}\right|_0 \right),
\label{RF_Pcoyy}
\\
   cP_{\rm co}^{zz} &\sim& 4\pi I_0 \frac{1}{3} \left( 1 + \left.\frac{12}{5}\frac{d\beta}{d\tau}\right|_0 \right).
\label{RF_Pcozz}
\end{eqnarray}
Using these results, the Eddington tensors become
\begin{eqnarray}
   f^{xx} &\sim& \frac{1}{3}
   \left( 1 - \left.\frac{8}{15}\frac{d\beta}{d\tau}\right|_0 \right),
\label{fxx}
\\
   f^{yy} &\sim& \frac{1}{3}
   \left( 1 - \left.\frac{8}{15}\frac{d\beta}{d\tau}\right|_0 \right),
\label{fyy}
\\
   f^{zz} &\sim& \frac{1}{3}
   \left( 1 + \left.\frac{16}{15}\frac{d\beta}{d\tau}\right|_0 \right),
\label{fzz}
\end{eqnarray}
which satisfy the relation $f^{xx}+f^{yy}+f^{zz}=1$.

Thus, the Eddington factor in the direction of the flow
slightly decreases,
whereas that in the direction perpendicular to the flow
slightly increases due to the velocity gradient.
Since we have assumed that
the emitted intensity is isotropic 
in the sufficiently optically thick regime,
there is no aberration effect,
and this result is due to the redshift effect,
where the one-tau photo-oval is seen to expand
for the comoving observer due to the velocity difference.

\section{Concluding Remarks}

In this paper, from the viewpoint of the comoving observer
we have analytically examined the Eddington factor
in the relativistic flow accelerating in the vertical direction;
we have introduced the one-tau photo-oval observed by the comoving observer,
and then calculated the comoving radiation fields and the Eddington factor.
We thus have shown that
the Eddington factor depends both on
the flow {\it velocity} and the {\it velocity gradient}
in the relativistic regime.

Under the present plane-parallel case,
the velocity increase in the downstream direction causes
the density decrease in the downstream direction;
this effect is essential for the shape of 
the one-tau photo-oval and vessel.
In the spherical flow, on the other hand,
there exists a geometrical dilution effect,
and the situation can be somewhat different.
Such a spherical case is another problem.

In this paper,
in order to analytically examine the relativistic variable Eddington factor
and explicitly demonstrate its dependence on the velocity gradient,
we restricted ourselves in the linear approximation,
where the velocity can be expanded,
and the velocity gradient is sufficiently small.
However,
the linear expansion of the velocity
may not be inadequate
when the velocity becomes large to be on the order of unity.
Hence, the present approximation would violate
in the region of $\beta \sim 1$.
We may examine such an extremely relativistic case
under the same approach.

Moreover,
in order to treat the problem in the simplest form,
we only consider the sufficiently optically thick regime,
where the one-tau photo-oval always exists.
In the optically thin regime or 
in the large velocity gradient case,
there may appear the photo-vessel,
where the optical depth becomes less than unity
in the downstream direction.
We should examine these general cases
in the future.

In the problem of line-driven winds with large velocity gradients,
the Sobolev approximation is often adopted
(e.g., Peraiah 2002; Castor 2004).
The Sobolev method was also extended to the relativistic regime
(e.g., Hutsem\'ekers, Surdej 1990; Hutsem\'ekers, Surdej 1993;
Jeffery 1993, 1995a, b).
Such a relativistic Sobolev approximation is also local
similar to the present case.
However, it is not so simple to compare
between the present continuum transfer case and 
the Sobolev's line transfer one.
Whether the present approach can be applied to the line transfer case,
or the comparison with the Sobolev method
is also the future work.

Although the present treatment is limited in the linear regime,
we first examined the relativistic variable Eddington factor
in the relativistic radiative flow under the physical situations,
and obtained its dependence on the velocity and the velocity gradient
in the explicit form.

The relativistic variable Eddington factor
for relativistic radiative flows discussed in the present paper
is fundamentally important
in various aspects of relativistic astrophysics with radiation transfer;
i.e., 
black-hole accretion flows with supercritical accretion rates,
relativistic jets and winds driven by luminous central objects,
relativistic explosions including gamma-ray bursts,
neutrino transfers in supernova explosions,
and various events occured in the proto universe.

\vspace*{1pc}

The author would like to thank M. Umemura,
T. Koizumi, and C. Akizuki
for enlightening and stimulated discussions.
This work has been supported in part
by a Grant-in-Aid for Scientific Research (18540240 J.F.) 
of the Ministry of Education, Culture, Sports, Science and Technology.


\end{document}